\input{aipcheck}

\documentclass[
    ,final            
  ]
  {aipproc}

\layoutstyle{8x11single}

\begin{document}

\title{How much complementarity?}

\author{Ingemar Bengtsson}{
  address={Fysikum, Stockholms Universitet, 106 91 Stockholm, Sweden}
}


\begin{abstract}
Bohr placed complementary bases at the mathematical centre point of 
his view of quantum mechanics. On the technical side then my question 
translates into that of classifying complex Hadamard matrices. Recent 
work (with Barros e S\'a) shows that the answer depends heavily on 
the prime number decomposition of the dimension of the Hilbert space. 
By implication so does the geometry of quantum state space.
%
%
\end{abstract}

\maketitle


\section{Reading Bohr}

Reading what Bohr actually wrote about the foundations of quantum mechanics 
one is struck by the modesty of his aims \cite{Bohr}. To Bohr, the aim of 
the theory is to predict the outcomes of measurements performed on a suitably 
prepared system. In a possibly double edged endorsement of Bohr's position, 
Mermin stresses how suitable he finds this view when teaching quantum mechanics 
to students coming from computer science: they want input and output, and 
have no emotional attachment to what goes on in between \cite{Mermin}. This 
clearly represents a retreat from the natural position of the physicists, who 
used to think that the essence of the phenomena resides there---and were then 
explicitly told by Bohr not to try to disclose them. If Bohr solved 
the interpretational problem of quantum mechanics then---as Marcus Appleby 
told me one fine day in front of the Rosetta stone---the problem 
is to find a point of view from which this solution appears desirable.

It is striking too how little of the mathematical formalism Bohr brings up. 
The one mathematical point stressed by him is the occurence, in quantum 
mechanics, of complementary pairs of measurements: if the system has been 
prepared to give a definite answer for one of them, nothing is known about 
the outcome should the complementary measurement be made \cite{Bohr}. Bohr's 
choice here shows good judgment. It may not be an ideal starting point for axiomatic 
reconstructions of the theory, but certainly the whole structure can be made 
to flow naturally through there---as Schwinger so convincingly demonstrated 
\cite{Schwinger}. So Bohr's vision cannot be dismissed lightly. 

At this point the discussion can go in many directions, philosophical and 
technical. The former may be more urgent \cite{Stig}, but my very modest aim 
here is to discuss how much freedom one has in choosing 
the complementary measurement. Because of the unitary symmetry 
the answer is independent of the choice of the first measurement, but it will 
turn out to depend in an interesting way on which Hilbert 
space we are in. 

\section{Complementary pairs of bases}

Let us assume that the dimension of our Hilbert space is $N$. We will have to 
come back to the question what this means. Meanwhile we associate measurements 
to orthonormal bases in the familiar way. If two such measurements are 
complementary it must be 
true that the pair of orthonormal bases 
$\{ |e_i\rangle \}_{i=0}^{N-1}$ and $\{ |f_i\rangle\}_{i=0}^{N-1}$
are related by

\begin{equation} |\langle e_i|f_j\rangle |^2 = \frac{1}{N} \end{equation}

\noindent for all the basis vectors. The question now arises whether 
complementary pairs of bases exist in every dimension, and if so how many 
such pairs exist, counting them up to the natural equivalence under unitary 
transformations \cite{Kraus}. 

This problem is equivalent to another that has been studied for a long time. 
Let us form a matrix with elements 

\begin{equation} H_{ij} = \langle e_i|f_j\rangle \ . \end{equation}

\noindent If the two bases are complementary this is a complex Hadamard matrix, 
that is a unitary matrix all of whose elements have the same modulus. In this 
way the existence of a complementary pair is equivalent to the existence of a 
complex Hadamard matrix. It is natural to use one of the members of the 
pair as our computational basis, in which case the columns of the Hadamard 
matrix are given by the elements of the vectors in the second basis. Of course 
we are not interested in the order or the overall phases of these vectors, so 
we will regard two unitary matrices $H'$ and $H$ as equivalent if there exists 
a permutation $P$ and a diagonal unitary $D$ such that 

\begin{equation} H' = HDP \ . \end{equation}

\noindent But there is still some freedom in the choice of the coordinate system. 
Given a pair of bases represented by the unit matrix and an Hadamard matrix $H$, 
an overall unitary transformation (from the left) with a permutation and a 
diagonal unitary can be undone from the right when it acts on the unit matrix, 
while any Hadamard matrix $H$ becomes a new Hadamard matrix $H'$. 
So in classifying pairs of complementary bases up to unitary transformations 
we will regard two complex Hadamard matrices as equivalent if there exist 
diagonal unitaries $D_1, D_2$ and permutation matrices $P_1,P_2$ such that 

\begin{equation} H' = P_1D_1HD_2P_2 \ . \label{equivalence} \end{equation}

\noindent If this is so we say that $H$ and $H'$ are equivalent, written $H \approx 
H'$  \cite{Haagerup}. The problem of classifying all complementary bases up to 
overall unitary transformations is equivalent to classifying all complex Hadamard 
matrices up to this equivalence. (Classifying all triples of mutually 
complementary bases is a more involved affair, since the freedom of multiplying 
from the left will be restricted.) We can remove some of the ambiguity by insisting 
that all Hadamard matrices should be presented in dephased form, meaning that 
all entries in the first row and the first column equal $1/\sqrt{N}$. 

A complex Hadamard matrix of any size exists. A solution is the Fourier matrix 
$F_N$, with entries that are roots of unity only: 

\begin{equation} F_{ij} = \frac{1}{\sqrt{N}}\omega^{ij} \ , \hspace{6mm} 
\omega = e^{\frac{2\pi i}
{N}} \ , \hspace{6mm} 0 \leq i,j < N - 1 \ . \end{equation}

\noindent And indeed this is a matrix with many applications. But are there 
other solutions? It is known that a generic unitary matrix is determined by 
the moduli of its matrix elements up to the $2N-1$ phases that are removed 
by dephasing \cite{Karabegov}, so if the answer is ``yes'' then Hadamard 
matrices are quite exceptional among the unitaries. 

Our question has a long history. In 1867 the British mathematician Sylvester 
gave many examples of 
such matrices \cite{Sylvester}. Sylvester also proved uniqueness for 
$N = 2,3$. In 1893 the French mathematician Hadamard studied the case 
$N = 4$ \cite{Hadamard}, and found that any Hadamard matrix of this size 
is equivalent to 

\begin{equation} H(z) = \frac{1}{2} \left( \begin{array}{rrrr} 1 & 1 & 1 & 1 \\ 
1 & z & -1 & - z \\ 1 & -1 & 1 & - 1 \\ 1 & - z & - 1 & 
z \end{array} \right) \approx 
\frac{1}{2} \left( \begin{array}{rrrr} 1 & 1 & 1 & 1 \\ 
1 & -1 & z & - z \\ 1 & 1 & - 1 & - 1 \\ 1 & - 1 & - z & 
z \end{array} \right)\ . \label{Had} \end{equation}
  
\noindent This is a one parameter family of dephased Hadamard matrices, 
since the phase factor $z$ is arbitrary and invariant (apart from its 
sign) under the transformations introduced in eq. (\ref{equivalence}). In 1997 
the Danish mathematician Haagerup proved that for $N = 5$ the Fourier 
matrix is again unique up to the natural equivalence \cite{Haagerup}. The 
$N = 6$ case is still open. An elegant family of dephased $N = 6$ Hadamard 
matrices with three real parameters was found by Karlsson \cite{Karlsson}, 
and there is strong evidence that a four parameter family should exist \cite{Skinner, 
Szollosi}. Perhaps it has Karlsson's 3-dimensional family as its boundary? 
An isolated example not belonging to any continuous family is also known 
\cite{Tao}. Finally there are many constructions available in higher dimensions, 
but we are not even close to a classification \cite{TZ}.    

The motivations behind these works were various. Sylvester's is a very joyful 
paper written when the notion of a matrix was new, and he was interested in 
all sorts of patterns he could observe in them. Haagerup's motivations 
stemmed from operator algebra. Other motivations for the study of complex 
Hadamard matrices come from quantum groups \cite{Banica}, and from various 
corners of quantum information theory \cite{Werner, Berge}. The work on 
$N = 6$ is largely inspired by the 
problem of Mutually Unbiased Bases in quantum theory, to which we will return. 
It is worth mentioning that the action of a complex Hadamard matrix can 
be implemented in the laboratory by means of linear optics \cite{Marek}, and 
that the $N = 6$ case is realistically within reach. 

It is intriguing that the answer to the question seems to depend so intricately 
on the dimension of Hilbert space, but one also wonders if it is at all 
possible to say something in general about a classification problem that 
moves this slowly. I will argue that one can, but first we should see what 
the existence of complementary pairs means for the geometry of quantum 
state space.

\section{The geometry of state space}

Mathematically a quantum state is represented by a density matrix $\rho$, 
that is an $N\times N$ complex matrix with non-negative eigenvalues. The 
set of all quantum states is a compact body of $N^2-1$ dimensions, with 
the pure states $\rho = |\psi \rangle \langle \psi|$ lying at its boundary. 
The distance $D$ between two density matrices is conveniently defined by 

\begin{equation} D^2(\rho_1, \rho_2) = \frac{1}{2}\mbox{Tr}(\rho_1-\rho_2)^2 
\ . \end{equation}

\noindent If we choose the maximally mixed state as the origin we can 
think of the container space as a real vector space, with the notion of distance 
coming from the scalar product

\begin{equation} \rho_1\cdot \rho_2 = \frac{1}{2}\mbox{Tr} (\rho_1 - \frac{1}{N} 
{\bf 1})(\rho_2 - \frac{1}{N}{\bf 1} ) \ . \end{equation} 

\noindent All the pure states lie on a sphere centered at the maximally mixed 
state. This sphere is called the outsphere, and the maximally mixed state will 
be chosen as the origin. An arbitrary state is formed as a mixture 
of pure states, and it follows that the set of all quantum states forms a 
convex body with an intricate shape.  The reason why the shape is intricate is 
that the symmetry group of the body is a small but continuous subgroup of the set 
of all rotations in $N^2-1$ dimensions---namely, if we ignore some discrete 
symmetries, the unitary group or more precisely the group $SU(N)/Z_N$. 
The pure states therefore form a small but continuous subset of the body's outsphere. 
We define the insphere as the largest 
sphere one can inscribe in the body. It is concentric with the outsphere, 
and the radius of the outsphere is $N-1$ times the radius of the insphere. 
(And we learn that the case $N = 2$ is special.) 

In order to get a feeling for what the shape is, one can ask what kind of 
regular polytopes, and similar understandable structures, one can inscribe 
in it. Indeed an orthonormal basis in Hilbert space corresponds to a regular simplex 
with $N$ vertices, inscribed in the body of all states, centred at the origin,  
and spanning a plane through the centre of dimension $N-1$. Every state $\rho$ 
lies in a simplex of this type. A complementary 
pair of bases spans two planes oriented with respect to each other in a 
special way. In fact the two planes are totally orthogonal, meaning that 
any vector in one of them is orthogonal to any vector in the other. Note 
that the totally orthogonal planes already point to the use complementary 
measurements have in quantum state tomography; other things being equal 
complementarity will minimize the uncertainties caused by the fact that, in 
a laboratory with access to a potentially infinite ensemble of identically 
prepared systems, only a finite number of measurement will actually be carried 
out \cite{Fields}. 

Once we have found a pair of simplices coming from a complementary pair, 
we can adjust the coordinates 
so that one basis is described by the computational basis, and then move 
the other simplex around by acting on the pair from the left with permutations 
and diagonal unitaries---operations whose action on the computational basis 
can be undone by irrelevant action from the right with the same type of 
unitaries. See eq. (\ref{equivalence}). So there is some freedom, but there 
will be more freedom the larger the set of inequivalent Hadamard matrices is 
found to be. The bottom line is that 
the existence of complementary pairs of bases is very much a question of 
the shape of the body of density matrices. 

We can go on in this way. Since 
$(N-1)(N+1) = N^2-1$ we can find $N+1$ totally orthogonal planes in 
our $(N^2-1)$-dimensional vector space. We can place a regular simplex of 
the appropriate size in each plane, but it is not at all clear that its 
corners correspond to pure states. By construction they lie on the outsphere, 
but they may well lie well outside the body 
of states, whose shape is so difficult to discern. But then again it may 
be possible to inscribe all these simplices in the body, in which case we 
say that we have a complete system of $N+1$ Mutually Unbiased Bases 
\cite{Fields}---and we have one more handle on the shape. 

After many trials in six dimensions \cite{Bruzda, BH, BS, Jam, RE}, 
most investigators are convinced that complete systems of MUBs exist if---this 
much is known \cite{Fields}---and only if---this is a conjecture only---the 
dimension of the Hilbert space is a power of a prime number. We will not be 
concerned with this problem here, but it does hang in the background. 
By the way the best known complete systems of MUBs \cite{Fields} can be obtained 
by choosing one special complex Hadamard matrix, and then multiplying it from the left by 
appropriate permutations and diagonal unitaries to construct the remaining 
$N-1$ complementary bases. I would be interested to know if this is true also 
for the more exotic examples that are known in some prime power dimensions 
\cite{Kantor}, but I don't. 

Of course the shape of the body of states can be studied in many other ways. 
But we are focussing on an important aspect of it.
  
\section{Families of Hadamard matrices}

Now let us consider the family of inequivalent Hadamard matrices given in 
eq. (\ref{Had}). By inspection we see that it includes the Fourier matrix 
(at $z = i$), but it also includes a real Hadamard matrix (at $z = 1$). The 
latter has an interesting form: it is the tensor product $F_2\otimes F_2$. 
On reflection we realise that whenever the dimension of Hilbert space is 
composite we can form Hadamard matrices from a pair of 
Hadamard matrices of size $N_1$ and $N_2$ in this way. But it will not always 
be true that $F_{N_1}\otimes F_{N_2}$ is inequivalent to $F_{N_1N_2}$. 
As a matter of fact they are equivalent if and only if 
$N_1$ and $N_2$ are relatively prime. This follows from some elementary 
group theory, because any Fourier matrix can be regarded as the character table of 
a cyclic group, and the cyclic group $Z_{N_1N_2}$ is isomorphic to the cyclic 
group $Z_{N_1}\times Z_{N_2}$ if and only if $N_1$ and $N_2$ are relatively prime. 
In prime power dimensions it is $F_p\otimes \dots \otimes F_p$, and not 
the inequivalent matrix $F_{p^k}$, that lays the golden eggs (i.e., a complete 
set of MUBs \cite{Fields}).  

There exists a construction due, in its most general form, to Di\c{t}\u{a} 
\cite{Dita}, allowing us to 
construct a continuous family in dimension $N = N_1N_2$ starting from one 
Hadamard matrix $H^{(0)}$ in dimension $N_1$ and $N_1$ possibly different 
Hadamard matrices $H^{(1)}, \dots , H^{(N_1)}$ in dimension $N_2$. It uses a 
warped tensor product. In dephased form  

\begin{equation} H = \left( \begin{array}{cccc} H^{(0)}_{0,0} H^{(1)} & H^{(0)}_{0,1}
D^{(1)}H^{(2)} & \dots & H^{(0)}_{0,N_1-1}D^{(N_1-1)}H^{(N_1)} \\ \vdots & \vdots & & \vdots \\ 
H^{(0)}_{N_1-1,0} H^{(1)} & H^{(0)}_{N_1-1,1}
D^{(1)}H^{(2)} & \dots & H^{(0)}_{N_1-1,N_1-1}D^{(N_1-1)}H^{(N_1)} \end{array} 
\right) \end{equation}

\noindent where $D^{(1)}, \dots , D^{(N_1-1)}$ are diagonal unitary matrices (with 
their first entries equal to one in order to obtain $H$ in dephased form). 
In this way the example of $N = 4$ generalises. 
It can be shown that the family arising from the Di\c{t}\u{a} construction using 
Fourier matrices as seeds interpolates between the non-equivalent matrices 
$F_{n^k}$ and $F_n\otimes \cdots \otimes F_n$ for all values of $n$ \cite{Nuno}. 
This somehow provides the beginning of a 
rationale for the existence of this family. 

Assuming that the parameters that may be present in the individual $H^{(i)}$ do 
not complicate matters, the intrinsic topology of these families---if we ignore 
some discrete equivalences, whose action has been completely worked out only in 
special cases \cite{Bruzda}---is that of a higher dimensional torus. They are 
examples of the more 
general class of affine families \cite{TZ}, in which all relations between the 
phases in the matrix are linear. But affine families are not the end of the story. 
For $N = 6$ we obtain affine families of at most 2 dimensions, while the set of all 
inequivalent Hadamard matrices has at least 3, and almost certainly 4, parameters. 
Moreover Karlsson's 3-dimensional family, which is known in explicit form, has a 
much more interesting geometry than the tori. Before all the discrete equivalences 
are taken into account it looks much like a circle bundle over a sphere, but with 
special points over (some copies of) the Fourier matrix, where the circles are 
blown up to tori. 

\section{All Hadamard matrices connected to the Fourier matrix}

To address the classification in general we first lower our aim a bit. Rather 
than ask for all complex Hadamard matrices, we ask for all smooth families 
of Hadamard matrices that include the Fourier matrix. This is really 
a question about the dimension of some algebraic variety. Following 
Fermi---``when in doubt, expand in a power series''---we attack it by 
multiplying the matrix elements in the Fourier matrix by arbitrary phase factors, 
which are then expanded in a series:

\begin{equation} F_{ij} \rightarrow F_{ij}e^{i\phi_{ij}} = F_{ij}\left( 1 + 
i\phi_{ij} - \frac{1}{2}\phi^2_{ij} + \dots \right) \ . \end{equation}

\noindent Then we try to solve the unitarity conditions order by order in 
the free phases $\phi_{ij}$, and count the number $d$ of free parameters 
that remain. To first order in the perturbation Tadej and 
\.Zyczkowski \cite{TZ} made this calculation. For dimension $N$ they 
found the answer

\begin{equation} D_1 = \sum_{n=0}^{N-1}\mbox{gcd}(n,N) \ , \label{defect} 
\end{equation}

\noindent where gcd$(n,N)$ denotes the greatest common divisor of $n$ and 
$N$. Subtracting $2N-1$, that is the number of trivial phases arising from 
eq. (\ref{equivalence}), this gives an upper bound on the number of free 
parameters in a smooth family of dephased Hadamard matrices containing the 
Fourier matrix. If $N$ is a prime this upper bound equals zero, 
so that we know that the Fourier matrix is isolated in the set of all 
Hadamard matrices. If $N$ is a power of a prime the upper bound is equal to 
the dimension of the family that arises from the Di\c{t}\u{a} construction, 
so that this family is the largest possible such family in this case. 
We tried to see what happens in the remaining cases.  

By now my question has become very technical, and for the details I have to refer 
to a paper by Barros e S\'a and myself \cite{Nuno}. In outline, our first step was to 
use the special properties of the Fourier matrix to write the equations 
in a more manageable form---in effect we calculate all bases complementary 
to the Fourier basis, rather than those complementary to the computational 
basis. To first order in the perturbation the equations 
are linear and homogeneous, and we recover eq. (\ref{defect}) in a very 
transparent way. To higher orders we still have to solve a linear system, 
but now with a heterogeneous part given by the lower order solution. To 
a given order $s$ these systems have solutions if and only if the lower order 
solutions obey consistency conditions which take the form of a set of 
multivariate polynomial equations of order $s$. If these conditions are 
non-trivial the true dimension drops below the first order result 
(\ref{defect}). Should this happen we have to solve the polynomial 
equations in order to determine by how much the dimension drops, and 
then we can proceed to higher orders ... 

Using a mixture of numerical and symbolical calculations we were able 
to carry through this program to quite 
high orders, for 24 different choices of $N$ not equal to a prime power. 
One case then stands out as being very special: $N = 6$, for which the 
consistency conditions hold trivially up to order 25 in the perturbation. 
At order 26 Mathematica quite reasonably refused to continue the calculation. 
Still, this gives considerable support to the conjecture that a 4-dimensional 
family of dephased Hadamard matrices does exist in this case---and it warns 
us not to make induction from six to arbitrary composite dimensions. 
$N = 10$ also stands out as somewhat special (the consistency conditions 
break down at order 11). In all other cases the consistency conditions 
break down in a systematic manner: at order 3 if $N$ is a product of three 
different primes, at order 4 if $N$ is a product of two different primes, 
at least one repeated, at order 5 if $N$ is a product of two odd primes, 
and at order 7 if $N$ is a twice an odd prime and larger than ten. 

We looked at the comparatively  manageable cases of $N = 12, 18, 20$ in more 
detail. We found solutions to the consistency conditions in symbolic form. 
For $N = 12$ we found the general solution to the consistency conditions 
at order 4, as well as an almost watertight argument saying that there 
does exist a two-sheeted solution such that no further breakdowns occur in 
higher orders. This was confirmed up to order 11 by an explicit 
calculation. Based on this information we conjecture that whenever 
$N = p_1p_2^2$, where $p_1, p_2$ are primes, there will be a non-linear 
family of dephased complex Hadamard matrices of dimension 

\begin{equation} d = 3p_1p_2^2 - 3p_1p_2 - 2p_2^2 + p_2 + 1 \ . 
\label{conj} \end{equation}

\noindent This number comes from the requirement that there should be two 
families, related by transposition and intersecting in a family arising 
from the Di\c{t}\u{a} construction in such a way that the two sheets 
span the whole tangent space---with its dimension given by 
the linearised calculation---when they intersect. 
We feel quite confident that this is true, but more to the point we feel 
that the mere fact that we were at all able to put forward a concrete 
conjecture suggests that there is a 
pattern here---we are not very close to a full solution of the problem 
for general $N$, but we do feel that we have a right to expect that eventually 
a solution will be found, in reasonably compact form.\footnote{Actually the 
argument in our published paper \cite{Nuno} is quite a bit stronger than the one I 
present here: here I tell the story as I knew it during the V\"axj\"o 
meeting.}   

That is to say, however odd the conjecture (\ref{conj}) may seem, we feel 
that it represents the beginning of a clear cut answer to the question 
posed in the title. 

\section{Conclusion}

A charge that has been raised against quantum mechanics is that of 
boring repetition: one might feel that the shape of the space of possible 
states should depend in an interesting way on the physical nature of the 
system, but in fact quantum mechanics uses 
the same old Hilbert space for everything \cite{Mielnik}. Perhaps what we 
have seen is a possible answer to this. The existence of pairs of 
complementary bases has an elegant interpretation in terms of the shape 
of the convex body of all possible states. And in this regard that shape 
does depend dramatically on the number theoretical properties of the 
dimension of the Hilbert space. 

But the dimension of the Hilbert space of a physical system is a property 
that can be measured, or at least be bounded from below by measurements 
\cite{Brunner, Wehner}. It has even been argued that the dimension 
of the Hilbert space is a candidate for the elusive role of something 
that goes on in between preparation and measurement \cite{Fuchs}. 
If we accept this, and if the results above have convinced us that 
the shape of the space of states does depend in an interesting way 
on the dimension of Hilbert space, then the charge against quantum 
mechanics falls. The shape and feel of the body of quantum states 
does depend on the physical nature of the system.  
 
\begin{theacknowledgments}
I thank Nuno Barros e S\'a for a wonderful collaboration, and Andrei 
Khrennikov for arranging such lively conferences. They 
comfort me for the fact that I was not able to attend the Warsaw 
conference---which was very lively too, judging from the proceedings 
\cite{Bohr}. 

\end{theacknowledgments}



\bibliographystyle{aipproc}   

\bibliography{sample}

\IfFileExists{\jobname.bbl}{}
 {\typeout{}
  \typeout{******************************************}
  \typeout{** Please run "bibtex \jobname" to optain}
  \typeout{** the bibliography and then re-run LaTeX}
  \typeout{** twice to fix the references!}
  \typeout{******************************************}
  \typeout{}
 }

\end{document}